# Resolving and Tuning Mechanical Anisotropy in Black Phosphorus via Nanomechanical Multimode Resonance Spectromicroscopy

Zenghui Wang[1,†], Hao Jia[1,†], Xu-Qian Zheng[1], Rui Yang[1], G. J. Ye[2],

X. H. Chen[2], Philip X.-L. Feng[1,*]

[1]*Department of Electrical Engineering & Computer Science, Case School of Engineering, Case Western Reserve University, 10900 Euclid Avenue, Cleveland, Ohio 44106, United States*

[2]*Hefei National Laboratory for Physical Science at the Microscale and Department of Physics, University of Science and Technology of China, Hefei, Anhui 230026, China*

*A*BSTRACT

**Black phosphorus (P) has emerged as a layered semiconductor with a unique crystal structure featuring corrugated atomic layers and strong in-plane anisotropy in its physical properties. Here, we demonstrate that the crystal orientation and mechanical anisotropy in free-standing black P thin layers can be precisely determined by spatially resolved multimode nanomechanical resonances. This offers a new means for resolving important crystal orientation and anisotropy in black P device platforms *in situ* beyond conventional optical and electrical calibration techniques. Furthermore, we show that electrostatic-gating-induced straining can continuously tune the mechanical anisotropic effects on multimode resonances in black P electromechanical devices. Combined with finite element modeling (FEM), we also determine the Young's moduli of multilayer black P to be 116.1 and 46.5 GPa in the zigzag and armchair directions, respectively.**

*K*EYWORDS:   *Black phosphorus, mechanical anisotropy, multimode resonances, spatial mapping, 2D materials, nanoelectromechanical systems (NEMS)*

[*]Corresponding Author.  Email:  philip.feng@case.edu.  [†]Equally contributed authors.

Efficiently exploiting anisotropic properties in crystals plays important roles in many areas in science and technology, ranging from timing and signal processing using a rich variety of crystalline cut orientations in quartz to the modulation and conversion of light using anisotropic crystals. In state-of-the-art miniature devices and integrated systems, crystalline anisotropy enables a number of important dynamic characteristics in microelectromechanical systems (MEMS) such as gyroscopes, rotation rate sensors, and accelerometers.[1,2,3,4] Single crystal silicon (Si), the hallmark of semiconductors and the most commonly used crystal in MEMS, possesses clear mechanical anisotropy that has been extensively characterized and utilized (*e.g.*, Young's moduli in the <110> and <100> directions are $E_{Y<110>}$ = 169 GPa and $E_{Y<100>}$ = 130 GPa).[5,6] As devices continue to be scaled down to nanoscale, anisotropy in mechanical properties may not always be readily preserved at device level due to lattice defects or surface effects,[7,8,9] and thus remain largely unexplored in emerging nanoelectromechanical systems (NEMS) built upon conventional crystals.

The recent advent of atomic-layer crystals offer exciting opportunities for building two-dimensional (2D) NEMS using single-crystal layered materials, in which many desired material properties are preserved, or even intensified, as the crystal thickness approaches genuinely atomic scale. One unique crystal is black phosphorus (P), not only a single-element direct-bandgap semiconductor with bandgap depending on the number of atomic layers (covering a wide range from visible light to IR) and with high carrier mobility, but also hitherto the best-known atomic-layer crystal with strong in-plane anisotropy. The intrinsically anisotropic lattice structure (Figure 1a) of black P dictates a number of anisotropic material properties. In particular, it is theoretically predicted to exhibit in-plane mechanical anisotropy (Figure 1a)[10,11,12,13] much stronger than that of Si, which shall lead to previously inaccessible dynamic responses in resonant NEMS[14] and new opportunities for studying carrier-lattice interaction in atomic layers.[15,16,17,18,19] To date, while extensive and rapidly growing efforts have been devoted to studying optical and electrical properties of black P and anisotropic effects in such devices, experiments on black P mechanical devices and mechanical anisotropic effects therein have been lacking. It is therefore of both fundamental and technological importance to systematically investigate the mechanical anisotropy in black P crystal.

In bulk materials, such as crystalline Si, mechanical anisotropy is often characterized by measuring the sound velocity in different directions,[20,21,22,23] and fitting data to the Christoffel equation.[24,25,26] Such conventional techniques, however, cannot be applied to micro and nanoscale structures because the sample dimensions (typically below 10 μm) become much shorter than the sound wavelength (~1 mm for 10 MHz ultrasound in Si), making it challenging to excite and measure sound wave propagating along well-defined directions. Currently, for characterizing mechanical properties in 2D nanostructures, the most common method is nanoindentation.[27] Nevertheless, for anisotropic crystal such as black P, prior knowledge of the crystal direction is required for fabricating specifically-orientated indentation samples that attempt to mechanically decouple the two crystal axes.[28] However, unlike in Si (where the crystal orientation is always indicated by the cuts on Si wafers with good precision),[5] such information is not readily available for 2D nanocrystals, and complete mechanical decoupling between the two axes has not yet been achieved.[28]

In this work, enabled by the first demonstration of black P resonant nanostructures with multimode responses, we show that the spatial mapping of the multimode resonance mode shapes creates a new means for the precise determination of black P crystal orientation (*i.e.*, the anisotropic zigzag and armchair axes). This is completely independent of conventional optical and electrical methods that require either polarized optical spectroscopic measurements [29,30,31,32,33,34] or device structures with many electrodes along multiple directions.[35,36] Furthermore, our technique enables simultaneous quantification of the anisotropic mechanical properties, *i.e.*, elastic moduli along both major crystal axes, without sophisticated device fabrication requiring predetermined crystal orientation.[28]

Black P resonant nanostructures are fabricated by employing an all-dry-transfer method with crystals synthesized in a high-temperature and high-pressure process.[37,38] Black P crystals are mechanically exfoliated onto a viscoelastic (polydimethylsiloxane, PDMS) stamp, and the candidate flakes are quickly identified and transferred (with alignment) onto microtrenches (depth of ~290 nm) prefabricated on a SiO$_2$–Si substrate with adjacent electrodes. This yields suspended, moveable nanomechanical devices with electrical contacts and tunability. In this work, we fabricate devices with both circular and rectangular shapes to demonstrate the capability of resolving mechanical anisotropy in black P crystal, as well as to investigate the interplay between the mechanical anisotropy and the device geometry (circle *vs.* rectangle, and orientation in the rectangular device).

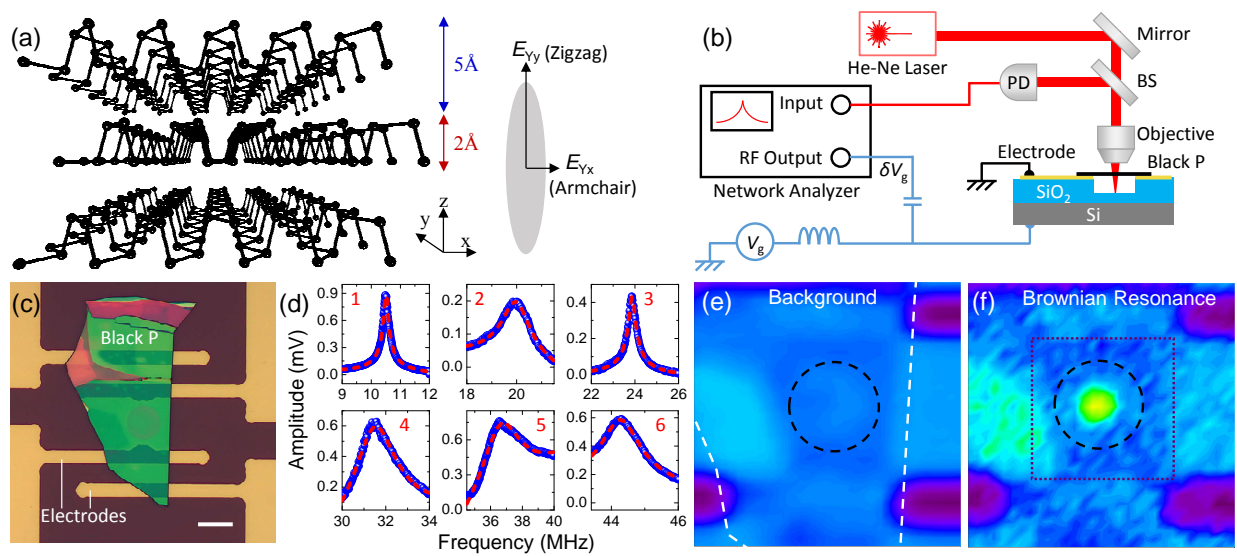

**Figure 1. Mechanical anisotropy in black phosphorus (P) crystal and multimode resonance characterization of black P nanoresonators**. (a) Schematic illustrations of black P crystal structure consisting of corrugated layers of P atoms and its intrinsically anisotropic elastic moduli along armchair ($E_{Yx}$) and zigzag ($E_{Yy}$) directions. (b) Custom-built measurement system, in which the out-of-plane motion of the black P nanoresonator is electrically excited by applying DC and AC gate voltages ($V_g+\delta V_g$), and interferometrically read out using a 633 nm He-Ne laser. Resonance modes are spatially-resolved by scanning spectromicroscopy measurements. (c) Optical image of a circular black P nanoresonator with electrodes. Scale bar: 10 μm. (d) Multimode resonance of the circular device in Figure 1c, showing six modes within 10 to 50 MHz. (e) Mapping of the off-resonance reflectance, showing uniform background across the device region. The black P flake and its suspended part are outlined. (f) Mapping of the fundamental resonant motion in the device's undriven Brownian motion, showing clear motion in agreement with the fundamental mode of a circular drumhead resonator. Green represents higher signal amplitude than does blue. The dashed circle indicates the suspended area. The dotted box indicates the mapping area in Figure 2a.

The multimode resonances of the black P nanoresonators are characterized using a scheme that incorporates electrical excitation and optical detection (Figure 1b): the out-of-plane vibrations in black P flakes are electrostatically excited by a back gate (except during calibration of the undriven Brownian resonances) and interferometrically detected by a He-Ne (633 nm) laser focused on the suspended flake.[38] The measurement techniques and the system have been engineered to achieve displacement sensitivities of ~36 fm/Hz$^{1/2}$ (calibrated for these black P devices) and submicron spatial resolution, and are thus capable of precisely mapping and vividly discerning the mode shapes of multimode resonators.[38,39]

Using this system, multimode resonances of black P nanoresonators are first characterized spectrally by performing microspectroscopy, harnessing frequency-domain multimode resonances for a given positioning of the readout laser spot.[39] Figure 1d depicts the 6 resonant modes observed in a circular device (Figure 1c, diameter $d$=10.0 μm, thickness $t$=95 nm) in the range from 10 to 50 MHz. The AC excitation voltages are 50 mV$_{rms}$ for Mode #1, 200 mV$_{rms}$ for Modes #2 and #3, and 1.2 V$_{rms}$ for Modes #4, #5 and #6 to accommodate the fact that at higher modes, the same driving force would lead to smaller displacements. These excitation voltages are also applied during the subsequent spectromicroscopy measurements. The mode shape of each resonance is then obtained by carefully performing the scanning spectromicroscopy measurements (focusing on each resonance mode, scanning the readout laser over the device surface and recording the spatial variations of the signal amplitude) to "map" the resonance mode shape and to attain its vivid 2D color plot for visualization.[39] We have also meticulously verified that the observed patterns truthfully represent the mode shapes of the multimode oscillations by characterizing off-resonance background (Figure 1e) and the fundamental resonant mode in undriven, Brownian motions (Figure 1f) of the same device. Mapping of driven resonances is performed over a 16 μm×16 μm area centered on the device (dashed box in Figure 1f), with 1 μm step size along both in-plane scanning axes.

We first examine the circular black P device in Figure 1c. The circular geometry (with the highest degree of symmetry among all 2D shapes) is suitable for revealing intrinsic material anisotropy, as the orientation-dependent device responses result only from the anisotropy in the crystal. Figure 2a shows the measured mode shapes for the six resonances in Figure 1d. The spatially resolved multimode resonances allow us to directly identify the black P crystal orientation and determine the two anisotropic axes. Circular black P diaphragms are expected to exhibit multimode resonances with a number of "mode pairs" (two modes with similar mode shapes that are rotated 90° from each other, *e.g.*, Mode #2 and #3 in Figure 2a).[14] Within each mode pair, the lower-frequency one has (in its mode shape) more nodal line(s) along the mechanically stiff direction (*i.e.*, zigzag axis in black P) and for the higher-frequency mode in the pair, along the soft (armchair) direction (Figure 2b, left side). In contrast, in an isotropic circular disk resonator the two modes within each pair would have the same nominal frequency (Figure 2b, right side). The mapping results in Figure 2a clearly show such signatures of mechanical anisotropy in the black P crystal: on the mapped mode shape of the lower-frequency Mode #2 (19.9 MHz), the dotted line highlights the nodal line that is aligned with the mechanically stiff zigzag direction, and on the mapped mode shape of the higher frequency Mode #3 (23.9 MHz), the dotted line indicates the resolved soft armchair direction. These spatial-mapping-resolved anisotropic crystal directions are depicted as red arrows in Figure 2c. Here, Mode #2 and #3 have a clear 20% difference from 19.9 MHz to 23.9 MHz. In sheer contrast, if consider the isotropic counterpart, the two modes within each degenerate pair would have the same nominal frequency (Figure 2b, right side) but with random, small splitting (*i.e.*, usually a ~1% to 0.1% difference with good fabrication precision), and the absolute orientation of nodal lines is random and uncontrollable.

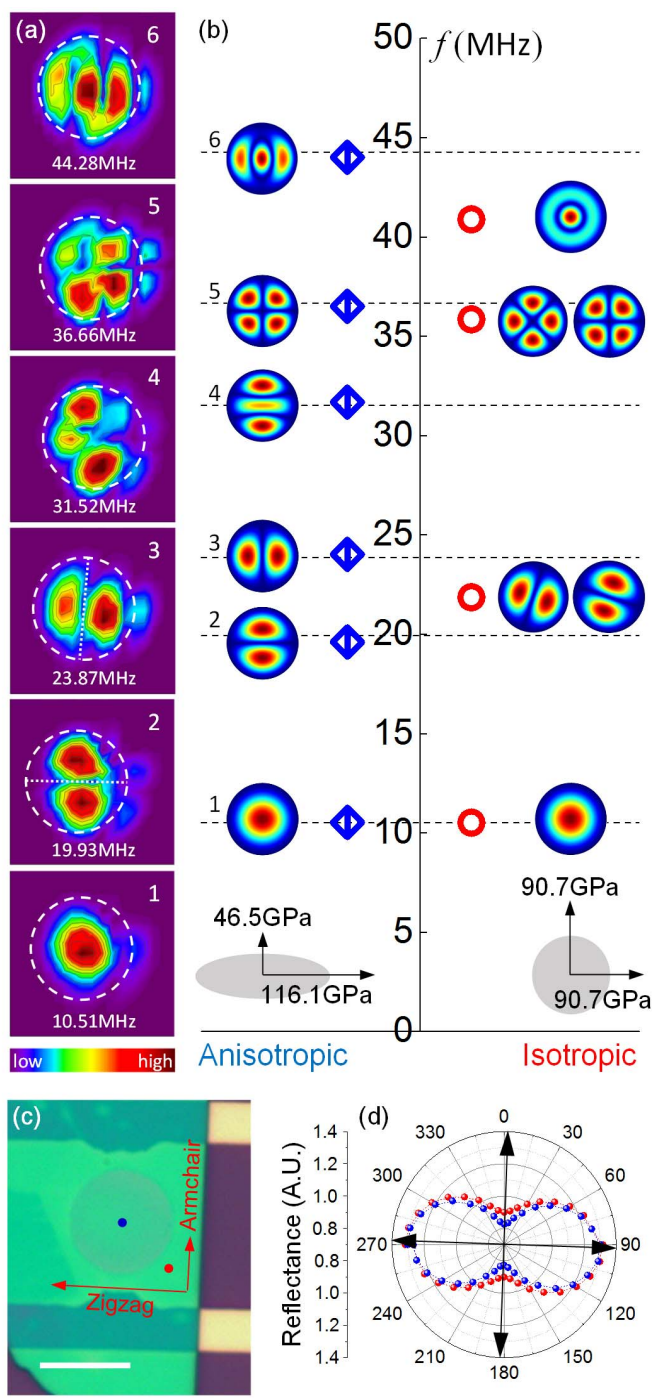

**Figure 2. Mechanical anisotropy in a circular black phosphorus (P) nanomechanical diaphragm determined by its spatially-resolved multimode resonances**. (a) Spatially-resolved flexural modes of the 10 μm circular device in Figure 1c. (b) Comparison of the measured resonance frequencies (dashed lines) with simulation results from an anisotropic model (blue diamonds, calculated using $E_{Yy}$= 116.1 GPa and $E_{Yx}$=46.5 Gpa) and an isotropic model (red circles, calculated using $E_Y$=90.7GPa which gives the same $f_{res}$ for the fundamental mode). The simulated mode shapes and schematic of the $E_Y$ values used are also shown for each model. (c) Optical image showing the crystal directions determined from the multimode resonance measurement. Scale bar: 10 μm. (d) Polarized optical reflectance measurement of the same flake using 532 nm illumination. The blue and red data are measured at the locations indicated by the blue and red dots in (c). The black arrows indicate the orientations of the two crystal axes extracted from the polarized optical measurement.

To validate the multimode resonance determination of crystal orientation, we further independently verify this by using polarized optical reflectance measurements.[34,28,40] The results (Figure 2d) show excellent agreement with the results in Figure 2c. This demonstrates that the spatially-resolved multimode resonances provide a new useful alternative for determining the crystal orientation, *in situ*, in the device platform of suspended black P nanostructures.

This nanomechanical resonance technique possesses several clear advantages that largely complement the existing methods for determining crystal orientation. First, this multimode resonance revelation method greatly exploits the effects of intrinsic mechanical anisotropy in the crystal and does not need repeated measurements at different laser polarization settings or complicated device structure with electrodes along many different directions, which are required in conventional methods. Second, our technique remains effective as the crystal thickness increases. While polarized optical measurements are widely used in characterizing ultrathin black P flakes,[29,30,31,32,33,34] as sample thickness increases, the crystal orientation associated with the higher Raman peak intensity (the same for higher-reflectance) alternates between zigzag and armchair directions (which further depends on probing wavelength),[31,41] making it challenging to discriminate the two crystal axes. In contrast, the mechanical anisotropy is well-preserved and manifested in the multimode resonant responses of multilayer samples[14] and can be effectively detected using the nanomechanical resonance technique. Third, the mode shapes allow for direct visualization of the crystal orientation (comparing the results from Modes #2 and #3), free from angle-based measurements in which the measurement configuration may limit the angular resolution in determining the crystal orientation (*e.g.*, in the most finely rotated polarized optical studies to date,[31] the sample is measured at every 10º, and in electrical measurements the angular resolution is limited by the number of electrodes that can be practically patterned, where the best angular resolution achieved so far is 30º).[35]

Not only do the multimode resonances provide deterministic information about the crystal orientation, but they also offer a quantitative and simultaneous measure of the elastic moduli of black P along both the zigzag and the armchair directions. To achieve this, we perform finite element modeling (FEM) simulations (by using COMSOL) for this circular device ($d$=10.0 μm, $t$=95 nm) using a model that accounts for the mechanical anisotropy and match the simulation results with measurements. Figure 2b summarizes the calculated resonance frequencies ($f_{res}$) from both the anisotropic model (blue diamonds) and an isotropic model (red circles, for comparison) and compares them with the measurements (horizontal dashed lines). We find excellent agreement with measurement for the anisotropic model, with a collective frequency mismatch of ~0.8% and a complete match between the mode sequences (determined from mode shapes), using $E_{Yy}$= 116.1 GPa (zigzag) and $E_{Yx}$= 46.5 GPa (armchair) (see more details in the Supporting Information). The measured mechanical anisotropy (elastic modulus ratio) is $E_{Yy}/E_{Yx}$=2.5, in good agreement with results from recent indentation measurements (~2.2).[28]

We now turn to devices in which the mechanical anisotropy in the black P crystal is coupled to the geometry (*i.e.*, shape and its orientation) of the device. One good example is a rectangular black P nanoresonator, in which the relative orientation between the rectangle's long side and the zigzag crystal axis (represented by the "orientation angle" $\theta$, Figure 3a) has strong effects upon the device's resonant responses.[14] Here, we demonstrate that one can quantitatively measure this orientation angle $\theta$ from the multimode resonant responses and thus precisely determine the crystal orientation in rectangular black P devices.

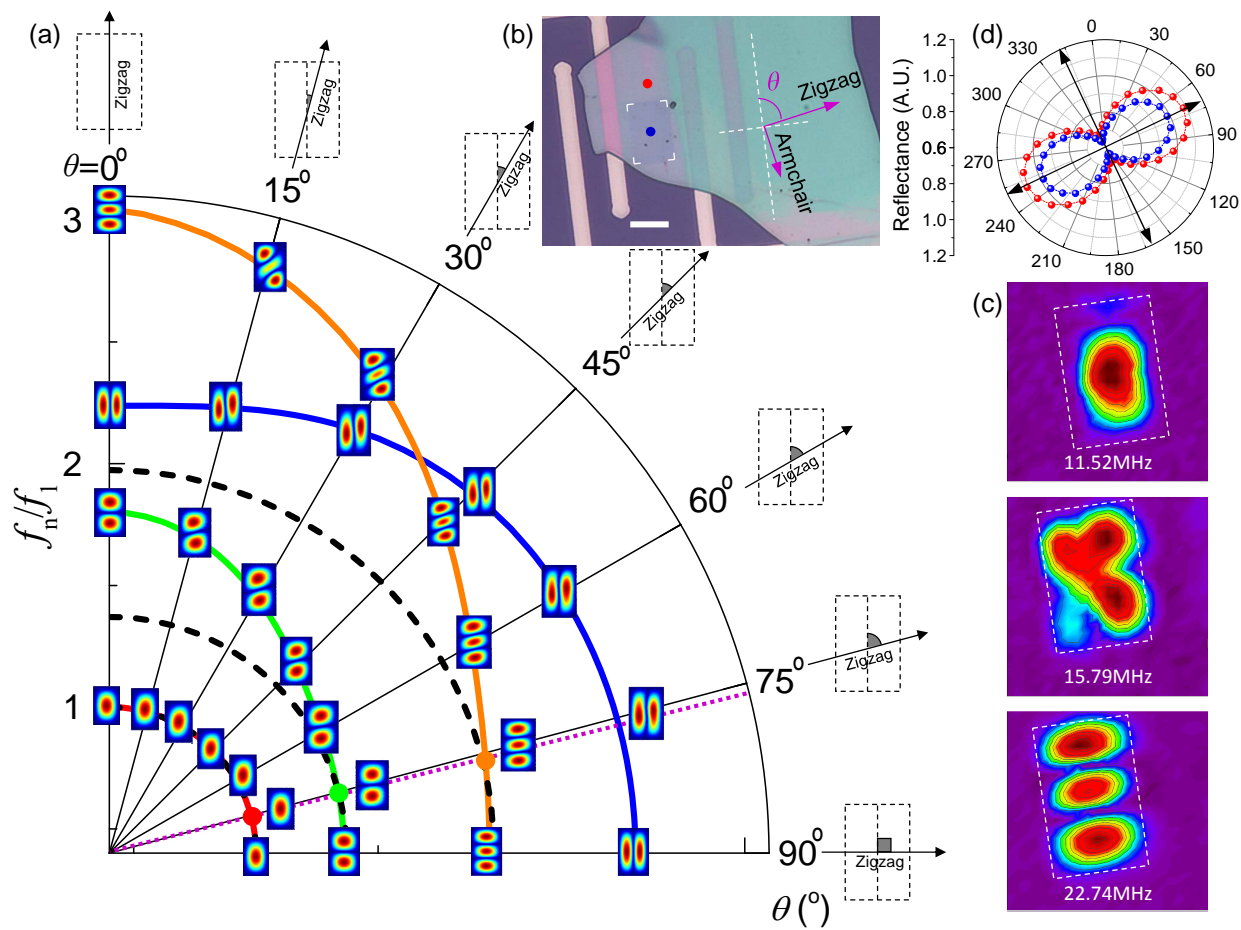

**Figure 3. Coupling between mechanical anisotropy and device geometry in black P nanoresonators.** (a) The polar plot shows the simulated multimode resonance of a rectangular black P nanoresonator with the orientation angle $\theta$ in the range of 0−90° (with mode shapes illustrated for every 15°). The magenta dotted line highlights the crystal orientation at which the calculation (colored solid curves) converges with the measurement (black dashed arcs). (b) Optical image of the device (the rectangular resonator is highlighted on all four corners). The rectangle orientation is indicated by the dashed cross. The crystal direction determined from the measurement is shown in magenta arrows, together with the resulting orientation angle $\theta$. Scale bar: 10 μm. (c) Measured mode shapes for the first three modes, showing good agreement with the simulation results at $\theta \approx 75°$. (d) Polarized optical reflectance measurement of the same flake using 532 nm illumination. Note that the 0° axis of the plot is chosen to be in alignment with the long side of the rectangle. The blue and red data are measured at the locations indicated by the blue and red dots in (b). The black arrows indicate the orientations of the two crystal axes extracted from the polarized optical measurement.

We experimentally study a rectangular black P resonator (Figure 3b, length $L$=15.1 μm, width $w$=9.9 μm, thickness $t$=186 nm) and measure its first three resonance modes (Fig. 3c shows the $f_{res}$ and mode shape for each mode. The AC excitation voltages are 0.1 $V_{rms}$ for Mode #1 and 1 $V_{rms}$ for Modes #2 and #3). Meanwhile, we perform FEM simulations for this device using the elastic moduli extracted from the aforementioned circular device and vary the orientation angle $\theta$ between 0º and 90º. The FEM calculated frequency ratios from the first four modes ($f_n/f_1$, n=1–4) are shown as colored solid curves (as functions of $\theta$) in Figure 3a, together with the simulated mode shapes. The simulation result clearly shows the coupling between the device geometry and mechanical anisotropy: for example, the 2$^{nd}$ lowest mode has the two antinodes strictly aligned along the rectangle's long side (rather than along the armchair direction as in circular devices) at both $\theta$=0º and $\theta$=90º, and for angles in between, the arrangement is tilted toward (but not strictly aligned to) the armchair axis. The coupling also leads to mode crossing between the 3$^{rd}$ and 4$^{th}$ modes at $\theta$≈38º: for $\theta$<38º, the 3$^{rd}$ mode has two antinodes aligned along the short side of the rectangle, and for $\theta$>38º it has three antinodes aligned along the long side.

To quantitatively determine the crystal orientation amid such coupled effects, in Figure 3a we compare the measured frequency ratios ($f_n/f_1$, dashed arcs) with simulation (solid curves) and find best agreement at $\theta$≈75º (magenta dotted line). The measured and simulated mode shapes also show good matching at this $\theta$ value. The crystal orientation determined using this method is illustrated in Figure 3b (magenta arrows). To independently verify the crystal orientation, we again perform polarized optical measurements (Figure 3d), and the results (black arrows in Figure 3d) show good agreement with the nanomechanical measurement (magenta arrows in Figure 3b). This shows that multimode resonant responses can effectively determine the crystal orientation in black P nanoresonators, even for rectangular-shaped devices in which the multimode resonances are determined by the coupling between the crystal's mechanical anisotropy and the device geometry.

Furthermore, we demonstrate electrical tuning of the mechanical anisotropy effects upon resonant responses. While the mechanical anisotropy in black P crystal offers new opportunities in engineering the resonance responses of black P devices, such as new mode sequences, frequency spacing, and mode shapes that are unavailable in isotropic devices, it is desirable that these resonant responses can be continuously tuned between the anisotropic and isotropic limits for attaining greater degrees of freedom in designing and controlling device response.

Here, we show that the electrical gate conveniently provides this tuning capability without requiring more-complicated device structures. Figure 4a shows the measured resonant responses of the circular black P resonator (the same device in Figures 1 and 2) over the gate voltage ($V_g$) range from $V_g$= -30 V to $V_g$ =30 V (the AC excitation voltages for all of these modes are kept at 10 mV$_{rms}$ to maintain linear operations of the device). The color represents signal amplitude (brighter color indicates higher amplitude). Each vertical slice is equivalent to an amplitude *vs*. frequency curve as in Figure 1d (except over a larger frequency range, thus including multiple resonances), and the entire plot represents measurements at different $V_g$ settings (see the Supporting Information for more details). We note that because this measurement is optimized for the first mode (laser spot at the device center), the peak amplitude for higher resonance modes are less visible on the same scale. Thus at each $V_g$, the resonance frequencies (with corresponding mode shapes indicated along $V_g$=0 V) are determined through fitting (as in Figure 1d) and denoted with color dots (one color for each mode). The mode shape of each resonance is indicated by the simulation results shown in the center and confirmed with mapping measurements. Here we focus on the two mode pairs besides the fundamental mode: the two-antinode pair (Modes #2 and #3), and the three-antinode pair (Modes #4 and #6), which best illustrate the transition from the anisotropic limit to the isotropic limit. From the data, we clearly observe both the capacitive softening effect ($f_{res}$ decreases as $V_g$ increases) which dominates at small $V_g$ values, and the elastic tensioning effect ($f_{res}$ increases with $V_g$) at higher $V_g$ values. We find that the frequency spacing between the two modes in each mode pair (*i.e.*, the "mode pair splitting", a signature of anisotropy in circular resonators; see Figure 2b) decreases, while $f_{res}$ for each individual mode increases with $V_g$. This indicates that the anisotropic effect is stronger at low $V_g$ values.

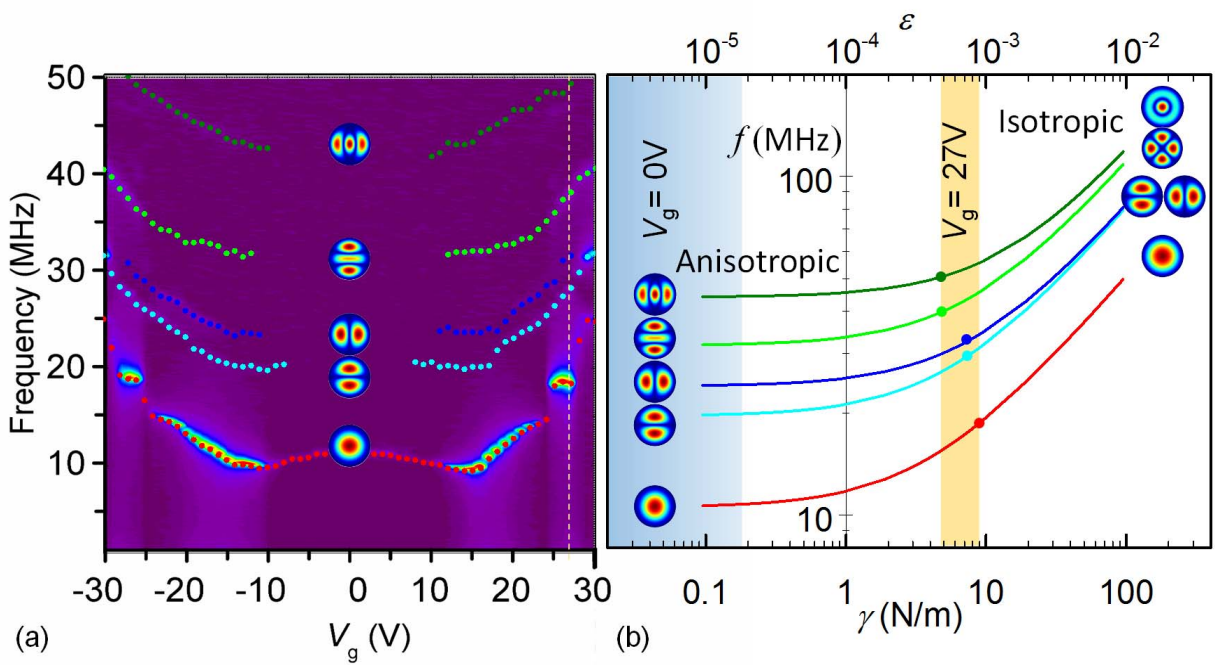

**Figure 4. Electrical tuning of effects from mechanical anisotropy in black P nanoresonator**. (a) 2D color plots showing multimode resonance of the 10 μm circular device in Figure 1c evolving with gate voltage (-30V≤$V_g$≤+30V). Brighter color indicates for higher signal amplitude. The mode shapes are determined from measurement shown in Figure 2 ($V_g$=0 V). The colored dots represent resonance frequencies extracted at each $V_g$ for the individual modes. (b) FEM simulation results from a model of a circular black P resonator under 2D biaxial tension ($\gamma$, in the unit of N/m as in surface tension), with the low tension limit $f_{res}$ values matching the $V_g$=0 V measurements (blue shaded area on the left). The evolution of $f_{res}$ (colored curves) clearly shows the transition of the device's multimode resonant response from an anisotropic limit to an isotropic limit when $\gamma$ varies from 0.1 to 100 N/m (corresponding to strain level $\varepsilon$=$10^{-5}$ to $10^{-2}$ in the zigzag direction). Simulated mode shapes are also shown for the two limits, again clearly showing the contrast between the two limits. Measured $f_{res}$ values (colored dots) at $V_g$=27 V (vertical dashed line in (a)) are compared to the calculation, suggesting that such gate tuning introduces a tension level of $\gamma$=4–8 N/m (orange shaded area).

To capture the key effect in such continuous tuning from anisotropy to isotropy, we use a simple model (via COMSOL) that focuses on the electrostatic tensioning (Figure 4b), in which we approximate the gating effect with a biaxial tension and examine the $f_{res}$ evolution for each mode. At $V_g$=0 V, the tension is negligible, and the $f_{res}$ values converge toward those from the FEM and the measurement results in Figure 2 (left-end blue shaded area in Figure 4b). As device tension ($\gamma$) increases, the simulation shows that the anisotropy-induced mode pair splitting decreases, and in the large tension limit, the resonant response of the black P resonator asymptotically approaches that of an isotropic device (in terms of both frequency spacing and mode shape, right-end in Figure 4b). The experimental data (Figure 4a) show a consistent trend, and as an example, we examine the multimode resonance at $V_g$=27 V (vertical dashed line in Figure 4a) by comparing the measured $f_{res}$ value for each mode (dots in Figure 4b) with simulation (curves in Figure 4b). We find good agreement between experiment and the simulation and further estimate a 2D tension on the order of ~10 N/m in this device at $V_g$=27V, which corresponds to ~0.1% of strain level (orange shaded area in Figure 4b). The simulation results suggest that at ~1% strain (right end of the curves in Figure 4b) the anisotropic effect would become negligible, and the entire device behaves as if it were made of an isotropic crystal. We note that this strain level is far below the fracturing strain limit of black P (~30%),[10] and although we have achieved ~0.1% strain by applying ~30 V gate voltage, a more effective straining mechanism may provide full-range tuning, including the switching "On" and "Off", of the mechanical anisotropy effects in the multimode responses of black P nanoresonators.

In conclusion, we have investigated the effects of mechanical anisotropy upon multimode responses in black P nanoresonators. We demonstrate that the multimode resonances (frequency, mode shape, and mode sequence) enable precise determination of the crystal orientation in these black P devices, independent of conventional angular-dependent optical or electrical measurements, for both circular and rectangular device structures. In addition, from the multimode resonant responses, we simultaneously determine the elastic modulus to be $E_{Yy}$= 116.1 GPa and $E_{Yx}$= 46.5 GPa (for zigzag and armchair directions, respectively) in the black P crystal. We further demonstrate that using electrostatic gating, the degree of anisotropy in the device's resonant responses can be continuously tuned, and at ~1% strain the device behavior will become mostly isotropic. Our results show that multimode resonances manifest the unique mechanical anisotropy effects in black P nanodevices and provide an independent method for determining the crystal orientation and elastic properties. Furthermore, we note that the applicability and effectiveness of the multimode resonance spatial mapping technique for discerning mechanical anisotropic properties in 2D crystals may be enhanced by combining it with insight into the device elastic characteristics. For the effective resolution and quantification of the anisotropic elastic moduli, the probing devices should be designed and made in their plate/disk regime (such as that demonstrated in this work. Note that being in this regime does not mean the device has to be thick; ultrathin but ultrasmall-diameter devices can be well in disk regime, *e.g.*, circular devices only 10 nm thick but with ≤2 μm diameter, and ≤0.1 N/m tension[14]). In conjugate, to responsively observe and quantify anisotropy in tension or biaxial stress, we shall design the probing devices to be in their membrane limit (note that being in this limit does not require a device to be thin, thick but fairly large and tensioned devices can exist well in the membrane limit, *e.g.*, 100 nm-thick circular devices, with ≥50 μm diameter, and ≥0.5 N/m tension). Together with the electrical tuning capability of mechanical anisotropy effects, multimode black P nanoresonators hold promises for new mechanically-active or flexible devices in which the material's mechanical anisotropy can be harnessed for enabling sensors, actuators, and dynamically-tunable electronic and optoelectronic transducers.

**SUPPORTING INFORMATION**

The supporting information is available free of charge on the ACS Publications website at http://pubs.acs.org.

Additional details on the extraction of elastic moduli of black P via multimode resonance spectromicroscopy, estimation of uncertainty in extracted Young's modulus, and resonance measurement parameters and electrical tuning of multimode resonances. Figures showing the extraction of elastic moduli in black P device by minimizing the "collective frequency mismatch" and measured multimode resonance curves (raw data). Tables showing mismatch between measured and simulated multimode resonances and frequency mismatch and $E_Y$ uncertainty for all six modes.

**ACKNOWLEDGMENTS**

We acknowledge the support from the Case School of Engineering, the National Academy of Engineering (NAE) Grainger Foundation Frontier of Engineering (FOE) Award (FOE2013-005), the National Science Foundation CAREER Award (ECCS #1454570), the CSC Fellowship (No. 201306250042 and 2011625071), and the CWRU Provost's ACES+ Advance Opportunity Award. Part of the device fabrication was performed at the Cornell Nanoscale Science and Technology Facility (CNF), a member of the National Nanotechnology Infrastructure Network (NNIN), supported by the National Science Foundation (ECCS-0335765). G. J. Y. and X. H. C. acknowledge the support from the National Natural Science Foundation of China.

# Resolving and Tuning Mechanical Anisotropy in Black Phosphorus via Nanomechanical Multimode Resonance Spectromicroscopy

Zenghui Wang[1†], Hao Jia[1†], Xu-Qian Zheng[1], Rui Yang[1], G. J. Ye[2],

X. H. Chen[2], Philip X.-L. Feng[1*]

[1]*Department of Electrical Engineering & Computer Science, Case School of Engineering, Case Western Reserve University, 10900 Euclid Avenue, Cleveland, OH 44106, USA*

[2]*Hefei National Laboratory for Physical Science at Microscale and Department of Physics, University of Science and Technology of China, Hefei, Anhui 230026, China*

[†]Equally Contributed Authors.  [*]Corresponding Author, Email: philip.feng@case.edu

**Extraction of Elastic Moduli of Black P via Multimode Resonance Spectromicroscopy**

As shown in Fig. 2 of the Main Text, we have clearly observed the effect of black P's intrinsic mechanical anisotropy on the device's multimode resonance characteristics. By taking advantage of such effect, we are able to not only unambiguously determine the black P's crystal orientation, but also quantitatively extract the elastic moduli of black P, *i.e.*, $E_{Yx}$ (along armchair direction) and $E_{Yy}$ (along zigzag direction) values.

The extraction process is performed by implementing a finite element method (FEM) model and matching the measured multimode resonances with the simulated ones: we first match their mode shapes (to ensure correct mode sequence), and then sweep the parameter in the model to obtain the best match in resonance frequency between measurement (dashed lines in Fig. 2) and simulation (blue diamonds in Fig. 2).

In the FEM model, a fully-clamped circular plate ($d$=10µm, $t$=95nm) is used and assigned with orthotropic material property to capture the in-plane mechanical anisotropy of black P crystal. The first 6 ($N$=6) eigenmodes are then calculated with a parametric sweep in the parameter space, $E_{Yx} \in$ (30–50GPa) and $E_{Yy} \in$ (100–130GPa). For each set of ($E_{Yx}$, $E_{Yy}$), we quantify the agreement between the simulated and measured resonance frequency values by computing the frequency "collective mismatch", which accounts for the fractional frequency mismatch for every mode:

$$\text{Collective Mismatch} = \sqrt{\frac{\sum_{i=1}^{N}\left(\frac{f_{i,\text{simulation}} - f_{i,measurement}}{f_{i,measurement}}\right)^2}{N}} \times 100\%, \qquad (S1)$$

where $f_{i,\text{simulation}}$ and $f_{i,\text{measurement}}$ are the simulated and measured frequency values of a particular resonance mode, $f_i$. Here fractional frequency mismatch ($\frac{f_{i,\text{simulation}} - f_{i,measurement}}{f_{i,measurement}}$), instead of the absolute frequency difference ($f_{i,\text{simulation}} - f_{i,measurement}$), is used in order to assign proper statistical weight to every mode regardless of its frequency value (otherwise higher modes would have higher weights).

We optimize $E_{Yy}$ and $E_{Yx}$ values by minimizing the collective mismatch. Figure S1 shows the collective mismatch as functions of $E_{Yy}$ and $E_{Yx}$. We find the minima (the black spot in Fig. S1a) on the colored surface when $E_{Yx}$=46.5GPa and $E_{Yy}$ = 116.1GPa, where the collective mismatch is only ~0.8%. Figure S1b shows the comparison between simulation and measurement for each individual mode and their corresponding mode shapes (detailed data shown in Table S1). In all, we have achieved excellent agreement among all the 6 modes, in terms of resonance frequency, mode sequence, and mode shape.

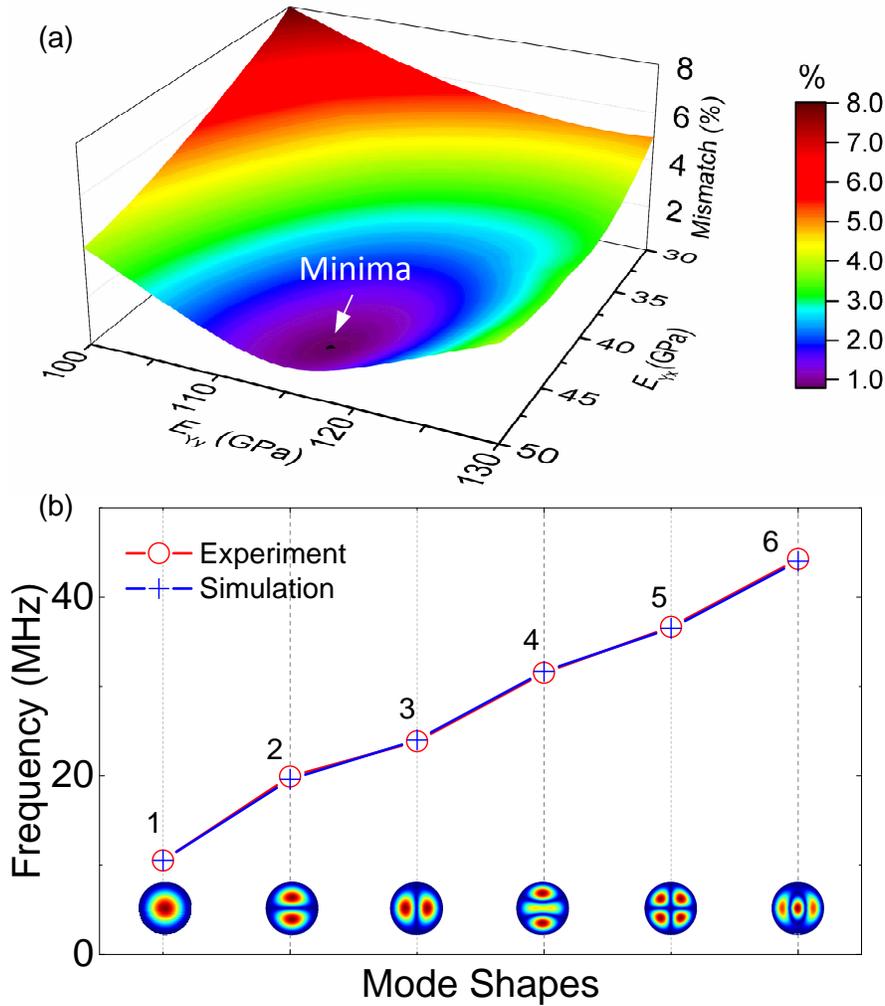

**Figure S1**: Extraction of elastic moduli in black P device by minimizing the "collective frequency mismatch". (a) 3D surface plot of the 'collective mismatch' as functions of $E_{Yx}$ and $E_{Yy}$ The minimal mismatch (~0.8%) is indicated by the black spot on the colored surface, located at $E_{Yx}$= 46.5 GPa, and $E_{Yy}$= 116.1GPa. (b) Excellent agreement between measured and simulated multimode resonances at this optimal point (where "collective mismatch" is minimized).

**Table S1**: Mismatch between Measured and Simulated Multimode Resonances at $E_{Yx}$=46.5GPa, $E_{Yy}$ = 116.1GPa

| Mode Shape | Measured Frequency (MHz) | Simulated Frequency (MHz) | Mismatch (MHz) | Mismatch |
|---|---|---|---|---|
| 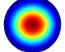 | 10.51 | 10.54 | 0.03 | 0.3% |
| 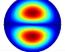 | 19.93 | 19.61 | -0.32 | -1.6% |
| 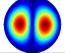 | 23.87 | 24.03 | 0.16 | 0.7% |
| 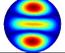 | 31.52 | 31.69 | 0.17 | 0.5% |
| 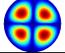 | 36.66 | 36.51 | -0.15 | -0.4% |
| 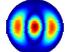 | 44.28 | 44.01 | -0.27 | -0.6% |

**Estimation of Uncertainty in Extracted Young's Modulus**

We estimate the uncertainty in the extracted elastic modulus from the frequency mismatch values (listed in Table S1). For each mode of a resonator operating in the plate regime, the elastic modulus is related to the resonance frequency *via*:

$$f_{res} \propto \sqrt{D} \propto \sqrt{E_Y}, \tag{S2}$$

where D is the bending rigidity. Note that this expression is true for resonators with any geometry (circular, rectangular, or others) and any boundary condition (supported or clamped, at center or perimeter). For resonators made of anisotropic materials (such as black P), this relation holds true for any given anisotropy (*i.e.*, elastic modulus ratio, $E_{Yy}/E_{Yx}$): for example, when the elastic moduli quadruples, the frequency of each mode doubles.

We can thus write down the expression for error propagation using Eq. S2, as:

$$\frac{\Delta f_{res}}{f_{res}} = \frac{1}{2}\frac{\Delta E_Y}{E_Y}. \tag{S3}$$

Equation S3 can be applied to each individual mode to estimate the error in $E_Y$ from the error in $f_{res}$. When multiple resonant modes are studied simultaneously, $f_{res}$ from N modes provide N estimations of the same $E_Y$. Increasing the sample size (*i.e.*, measuring more modes) also statistically reduces the uncertainty.

In this work, N=6 modes are measured. Counting all the 6 modes we have studied for the circular device, the collective uncertainty in estimating $E_Y$ is (by combining Eq. S2 and S3) given by:

$$\text{Collective Uncertainty} = \sqrt{\frac{\sum_{i=1}^{N}\left(2\frac{\Delta f_i}{f_i}\right)^2}{N}} = 2\sqrt{\frac{\sum_{i=1}^{N}\left(\frac{f_{i,\text{simulation}} - f_{i,\text{measurement}}}{f_{i,\text{measurement}}}\right)^2}{N}}. \tag{S4}$$

Using Eq. S4, we thus estimate an overall statistical uncertainty of 1.6% in the extraction of elastic moduli, with uncertainty contribution from each individual mode listed in Table S2.

**Table S2**: Frequency Mismatch and $E_Y$ Uncertainty for All 6 Modes

| Mode Shape | Frequency Mismatch (%) | $E_Y$ Uncertainty (%) |
|---|---|---|
| 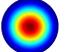 | 0.3% | 0.6% |
| 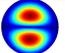 | -1.6% | -3.2% |
| 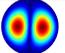 | 0.7% | 1.4% |
| 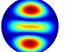 | 0.5% | 1.0% |
| 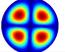 | 0.4% | 0.8% |
| 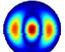 | 0.6% | 1.2% |
| Collective Values | 0.8% | 1.6% |

## Resonance Measurement Parameters and Electrical Tuning of Multimode Resonances

In the multimode resonance The AC voltages for the electrical drive (during mapping) are 50mV$_{rms}$ for Mode #1, 200mV$_{rms}$ for Mode #2 and 3, 1.2V$_{rms}$ for Mode #4, 5 and 6. These values are chosen to accommodate the fact that at higher modes, the same driving force would lead to smaller displacements, so as to attain good signal-to-noise ratio in spatially mapping the higher modes.

We have demonstrated gate tuning of the multimode resonances, as well as gate tuning of the effects of mechanical anisotropy in the Main Text (see Fig. 4). To further help illustrate the measured gate tuning effects, in Figure S2 we show a 3D view of the raw experimental data with the vertical axis being the signal amplitude (Fig. S2a). We have clearly observed the evolutions of multiple modes as the gate voltage ($V_g$) varies from -30V to +30V.

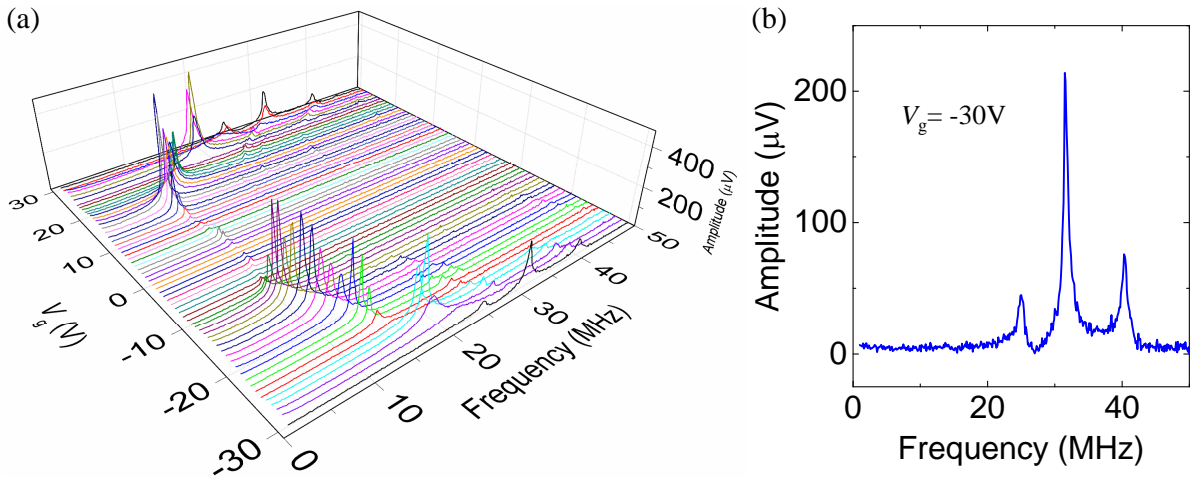

**Figure S2**: Measured multimode resonance curves (raw data) that have turned into the color-map plot in Figure 4a in the Main Text. (a) A 3D view of measured gate tuning of the multimode resonances, for the 10μm-diameter circular black phosphorus drumhead resonator under $V_g$ = -30V to +30V. (b) Simple line plot of measured multimode resonance response when $V_g$= -30V (*i.e.*, the rightmost curve in panel (a)).

In the range of $V_g$=-30V to +30V, the AC excitation voltage is fixed at ~10mV$_{rms}$, for all the curves shown in Fig. S2a. As one example, Figure S2b displays the ordinary line plot for one of the measured multimode resonance response (in the range of 1–50MHz) at $V_g$= -30V (*i.e.*, this is simply the rightmost curve in Fig. S2a).

In all the measurements in this work, we have limited the $V_g$ up to ±30V, and carefully selected the AC voltage such that the devices would behave in linear regime, to attain reliable and accurate results in gate tuning and spatial mapping measurements.

We also note that since this measurement is optimized for the first mode (laser spot positioned at the center of the device), the peak amplitudes of higher resonance modes are less visible on the same scale, as compared to the peak amplitude of the first mode. Thus at each $V_g$, the resonance frequency values are extracted through fitting to the damped simple harmonic oscillator equation, which are the data points represented by the colored, solid circles in Fig. 4a (one color for one mode) in the Main Text.